\documentclass{jfm}
\usepackage{graphicx}
\usepackage{natbib}
\usepackage{amsfonts}
\usepackage{amsmath}
\usepackage{graphicx,color}
\usepackage{dcolumn}
\usepackage{bm}
\usepackage{array}
\usepackage{booktabs}% for top-,mid- & bottomrule
\usepackage{caption}
\usepackage{color}

\ifCUPmtlplainloaded \else
  \checkfont{eurm10}
  \iffontfound
    \IfFileExists{upmath.sty}
      {\typeout{^^JFound AMS Euler Roman fonts on the system,
                   using the 'upmath' package.^^J}%
       \usepackage{upmath}}
      {\typeout{^^JFound AMS Euler Roman fonts on the system, but you
                   dont seem to have the}%
       \typeout{'upmath' package installed. JFM.cls can take advantage
                 of these fonts,^^Jif you use 'upmath' package.^^J}%
      }
  \else
  \fi
\fi
\ifCUPmtlplainloaded \else
  \checkfont{msam10}
  \iffontfound
    \IfFileExists{amssymb.sty}
      {\typeout{^^JFound AMS Symbol fonts on the system, using the
                'amssymb' package.^^J}%
       \usepackage{amssymb}%
         
         \let\geq=\geqslant
      }{}
  \fi
\fi
\ifCUPmtlplainloaded \else
  \IfFileExists{amsbsy.sty}
    {\typeout{^^JFound the 'amsbsy' package on the system, using it.^^J}%
     \usepackage{amsbsy}}
    {}
\fi
\def\d{{\rm d}}
\newsavebox{\astrutbox}
\sbox{\astrutbox}{\rule[-5pt]{0pt}{20pt}}

\def\Pe{{\rm Pe}}

\title[Stochastic dynamics of  active swimmers  in linear flows]{Stochastic dynamics of  active swimmers \\ in linear flows}

 \author[Mario  Sandoval, Navaneeth K.M.,  Ganesh Subramanian, and Eric Lauga]
{Mario Sandoval$^{1}$\thanks{Email address: sem@xanum.uam.mx},\ns
{Navaneeth K.~M.$^2$, Ganesh Subramanian$^2$},\ns  and
Eric Lauga$^{3}$\thanks{Email address: e.lauga@damtp.cam.ac.uk}}
\affiliation{
$^1$Department of Physics, Universidad Autonoma Metropolitana-Iztapalapa, Apartado Postal 55-534, Mexico, Distrito Federal 09340, Mexico.
\\[\affilskip]
$^2$Engineering Mechanics Unit, JNCASR, Bangalore 560064, India.
\\[\affilskip]
$^3$Department of Applied Mathematics and Theoretical Physics, 
University of Cambridge, Centre for Mathematical Sciences, Wilberforce Road,
Cambridge, CB3 0WA, United Kingdom.}

\begin{document}

\maketitle

\begin{abstract}
Most classical work on the hydrodynamics of low-Reynolds-number swimming  addresses  deterministic locomotion in quiescent environments. Thermal fluctuations in  fluids are known to  lead to a Brownian loss of the  swimming direction, resulting in  a transition from short-time ballistic dynamics to effective long-time diffusion. As most cells or synthetic swimmers are immersed in external flows, we consider theoretically in this paper the stochastic dynamics of a model active particle (a self-propelled sphere) in a steady general linear flow. The stochasticity arises both from translational diffusion in physical space, and from a combination of rotary diffusion and so-called run-and-tumble dynamics in orientation space. The latter process characterizes the manner in which the orientation of many bacteria decorrelates during their swimming motion. In contrast to rotary diffusion, the decorrelation occurs by means of large and impulsive jumps in orientation (tumbles) governed by a Poisson process. We begin by deriving a general formulation for all components of the long-time mean square displacement tensor for a swimmer with a time-dependent swimming velocity and whose orientation decorrelates due to rotary diffusion alone. This general framework is applied to obtain the convectively enhanced mean-squared displacements of a steadily-swimming particle in three canonical linear flows (extension, simple shear, and solid-body rotation). We then show how to extend our results to the case where the swimmer orientation also decorrelates on account of run-and-tumble dynamics. Self-propulsion in general leads to the same long-time temporal scalings as for  passive  particles in linear flows but with increased coefficients. In the particular case of  solid-body rotation, the effective long-time diffusion  is the same as that in a quiescent fluid, and we clarify the lack of flow-dependence by briefly examining the dynamics in elliptic linear flows. By comparing the new active terms with those obtained for passive particles we see  that swimming can lead to an enhancement of the mean-square displacements by orders of magnitude, and could be relevant for   biological  organisms or synthetic swimming devices  in fluctuating environmental or biological flows.

\end{abstract}


\section{Introduction}

A complete physical understanding of many processes occurring at small scales and involving active particles has proven both challenging and an exciting avenue for biomechanics and bioengineering research. Important biological topics with ongoing  research include  the dynamics of plankton in marine ecosystems \citep{guasto12},  the collective behaviour of dense micro-organism suspensions \citep{koch} and their appendages \citep{lg12}, and the interactions between swimming cells and complex environments \citep{lp09}. In the bioengineering world, the focus is on the design of  effective, and practical synthetic locomotion systems able to carry out future  detection, diagnosis, and treatment of diseases 
\citep{paxton06,chemical2,chemical3,abbot}.

Focusing on the dynamics of a single active particle or self-propelled  cell, most classical work considered the kinematics and energetics of deterministic locomotion in a quiescent fluid. Due to their small sizes, many swimming cells, in particular, bacteria and small single-cell eukaryotes, as well as many synthetic swimmers, are expected to have their swimming direction affected by thermal fluctuations \citep{lovely75,ped,berg93,ishi,lauga,howse07,hagen,hage2}. Even for bacteria  large enough to not be Brownian, there continue to be stochastic fluctuations in orientation that are largely athermal in origin. For instance, for a bacterium \textit{E.~coli} during a swimming run,  the observed rate of orientation decorrelation is one order of magnitude faster than that predicted based on a rotary Brownian diffusivity \citep{berg93}, and is likely due to  shape fluctuations of the imperfect  bundle of bacterial flagella  \citep{SubKoch,koch,locsei2009run}.

Furthermore, in most situations of biological or applied interest, self-propelled organisms and synthetic swimmers are subject to external flows, for example plankton transported by  small-scale turbulence, bacteria in the initial stages of environmental biofilm formation or swimming through human organs. Similarly, any future practical implementation of artificial micron-scale swimmers will have to be able to navigate through flowing bodily fluids, in particular the bloodstream \citep{abbot,kosa,wang2012}.

Previous classical studies have addressed the effect of external flows on Brownian motion of passive spherical colloids, most notably in simple shear \citep{miguel,SubBrady} and for the more general case of an arbitrary incompressible linear flow  \citep{foister}.
For a passive spherical Brownian particle in a linear flow, the long-time diagonal element of its mean-square displacement dyadic along the flow direction is proportional to the third power of time in the case of simple shear, grows at an exponential rate (along the extensional axis) in the case of pure extensional flow, but continues to display a diffusive scaling in the case of solid-body rotation. In the case of a shear flow, \cite{clercks} studied a similar problem  based on the time dependent linearized Navier-Stokes equations, instead of the Stokes equations, in order to address the non-trivial modifications in the short-time dynamics arising from the inclusion of fluid inertial effects. The analogous situation in the absence of a flow is  classical work \citep{hinch1975,zwanzig1970,hauge1973}, and accounting for the finite time scale on which vorticity diffuses leads to an algebraic (rather than exponential) decay in the relevant correlations. For non-spherical particles, the dynamics cannot in general be obtained in closed form, since the translational dispersion is intimately coupled to the orientation distribution, and the latter cannot be determined analytically for arbitrary values of the rotary P\'eclet number  \citep{brenner1991,brenner1993}. Asymptotic analysis is, however, possible both for small values of the rotary P\'eclet number \citep{brenncond1974,brenner1974} and in the limit of weak Brownian motion \citep{leal}

The dynamics of active particles  in shearing flows has been addressed in recent studies.  \cite{jones} calculated, in the absence of noise, the direction of swimming of bottom-heavy micro-organisms immersed in shear flows.  \cite{bearon} modelled a spherical chemotactic {bacterium}  and derived an  advection-diffusion equation for the cell density which included the influence of shear.   \cite{locsei2009run} addressed the run-and-tumble dynamics of bacteria and the effect of a shear flow on the chemotaxis response of the cell. 
 More recently, \cite{hagen2} characterised, in two spatial dimensions, the dynamics of a spherical self-propelled particle in a shear flow and subject to an external torque, and obtained an enhancement of the $\sim t^3$ mean-square dynamics. The effect of an external linear flow on the rheology of, and the pattern formation by, suspensions of active particles was   considered by \cite{saintillan_ext}, \cite{saintillan}, \cite{rafai} and \cite{pahlavan}.

In this paper we  quantify the interplay between fluctuations (thermal or otherwise) and a  prototypical external flow --  namely a steady, incompressible  linear flow --   on the dynamics of an active particle.  The particle is assumed to be spherical, a geometry relevant to many biological and bioengineering situations, including the dynamics of self-catalytic colloidal spheres \citep{howse07,golestanian07,BRADY:2010bh,Julicher:2009hw}, active droplets \citep{Thutupalli:2011bv,Schmitt:2012ts}, and the algae {\it Volvox} \citep{drescher09}. The activity of the  particle, which is free to move in three spatial  dimensions, is  modelled as a prescribed swimming velocity in its body frame. We first develop the analysis in the case where the particle is subject to both rotational and translational Brownian motion, in addition to being convected by the ambient linear flow. We then extend the results   to include the biologically relevant re-orientation mechanism associated with the run-and-tumble dynamics exhibited by many  bacteria \citep{berg93,bergbook}. We ignore other potentially relevant reorientation mechanisms, including phase slips which occur between the pair of anterior flagella of the { {\it Chlamydomonas}} algae \citep{Polin2009}, hydrodynamically mediated collisions that govern the dynamics at high volume fraction \citep{ishi}, and run-and-reverse dynamics \citep{guasto12}.

After setting up the problem in \S\ref{setup}, we derive in \S \ref{S:rot}, by means of an elementary rotational transformation, the transition probability density for a particle whose orientation evolves on account of a rotary diffusion process. We then use this probability density to find all components of the swimming direction correlation matrix. We exploit these results to calculate the  general expression for the mean-square displacement dyadic of the active particle in \S\ref{general} and evaluate each of its components analytically in the specific case of an active particle undergoing steady swimming in \S\ref{stead}.  In the absence of external flow, or for passive particles, our analytical results recover the well-known classical limits. By focusing on three prototypical flows (simple shear, extension and solid-body rotation) in \S\ref{3cases}, we demonstrate that the particle activity does not modify the long-term temporal scalings for the mean-square displacements, but increases its coefficients in the case of shear and extension while the results are unchanged in the case of solid-body rotation. In \S\ref{Section_tumbling}, we extend the analysis to include an additional intrinsic orientation de-correlation mechanism, namely that associated with correlated tumbles, the  occurrence of which is modelled as a Poisson process. We demonstrate that   the effect of tumbles may be simply incorporated as an additive contribution to the rate of orientation de-correlation, and the results already obtained may therefore be readily extended to include swimmers whose orientation evolves due to  both rotary diffusion and run-and-tumble dynamics. We close by offering a  physical discussion of our results  in \S\ref{discussion} using  scaling arguments. In particular, we explain the singular flow-independent nature of solid-body rotation by considering  the behaviour of the mean-squared displacement in elliptic linear flows (i.e.~two-dimensional linear flows with closed streamlines), and examining its dependence on the ratio of the ambient vorticity to extension. Comparing the coefficients in the active vs.~passive case, we see that swimming can lead to enhancement of the mean-square dynamics by orders of magnitude, a result which could be relevant for both biology and bioengineering.

%%%%%%%%%%%%%%%%%%%%%%%%%
\section{A spherical active particle in an incompressible linear flow}
\label{setup}

We consider a spherical particle of radius $a$ that self-propels (swims) in a three-dimensional fluctuating environment and in the presence of a general linear external flow. In the absence of noise and external flow, we assume that the particle swims at the intrinsic velocity $\mathbf{U}_{s}(t)$, prescribed in the body frame of the particle. We
use a cartesian coordinate system with vectors $\{\mathbf{i,j,k\}}$ and
corresponding coordinates $(x_{1},x_{2},x_{3})$. The external flow, $\mathbf{U}_{\infty }$, is assumed to be any general two-dimensional linear, incompressible flow of the form $\mathbf{U}_{\infty }=(Gx_{2},\alpha Gx_{1},0)$, with $G>0$ denoting the deformation rate. The
particle orientation is described by the angles $(\theta ,\varphi )$ in a
spherical coordinates system, where $\theta $ and $\varphi $ are the polar
and azimuthal angles respectively. The dimensionless parameter $\alpha$ { allows us }to tune the type of external flow considered, from pure rotation ($\alpha=-1$) to shear ($\alpha=0$) and extensional flow ($\alpha=1$).

The over-damped balance of forces and torques on the particle leads to the
Brownian Dynamics equations determining its instantaneous translational
velocity, $\mathbf{U}(t)$, and angular velocity, $\mathbf{\Omega }(t)$, as
solutions to 
\begin{equation}
\mathbf{R}_{U}(\mathbf{U}-\mathbf{U}_{s}-\mathbf{U}_{\infty })=\widetilde{\mathbf{f}},\quad \mathbf{R}_{\Omega }\left( \mathbf{\Omega }-\mathbf{\Omega 
}_{\infty }\right) =\widetilde{\mathbf{g}},  \label{din}
\end{equation}
where $\mathbf{\Omega }_{\infty }=\omega _{\alpha }\mathbf{k}$, with 
$\omega _{\alpha }=(G/2)(\alpha -1)$, is the
angular velocity of the particle induced by the general linear flow. In
equation \eqref{din}, $\mathbf{R}_{U}=R_{U}\mathbf{I}$ and $\mathbf{R}%
_{\Omega }=R_{\Omega }\mathbf{I}$ are the viscous resistance coefficients ($%
R_{U}=6\pi \eta a$ and $R_{\Omega }=8\pi \eta a^{3}$ in a Newtonian fluid of
shear viscosity $\eta $) and $\mathbf{I}$ is the unit tensor. The vectors $%
\widetilde{\mathbf{f}}$ and $\widetilde{\mathbf{g}}$ represent zero-mean
Brownian random forces and torques whose correlations in their components
are governed by the fluctuation-dissipation theorem as
\begin{equation}
\left\langle \label{equationbrownian}
\widetilde{{f}}_{i}(t)\widetilde{{f}_{j}}(t^{\prime })\right\rangle
=2k_{B}TR_{U}\delta _{ij}\delta (t-t^{\prime }),\,\,\left\langle 
\widetilde{{g}_{i}}(t)\widetilde{{g}_{j}}(t^{\prime })\right\rangle
=2k_{B}TR_{\Omega }\delta _{ij}\delta (t-t^{\prime }),
\end{equation}
with $\left\langle
\cdot \right\rangle $ representing ensemble averaging \citep{edwards}.

Denoting the particle location as $\mathbf{x}\left( t\right)
=(x_{1}(t),x_{2}(t),x_{3}(t))^T$, the equation governing $\mathbf{x}\left( t\right)$, from equation~(\ref{din}), can be formally written as 
\begin{equation}
\frac{\mathrm{d}{\mathbf{x}}}{\mathrm{d}t}=\mathbf{Mx}(t)+U_{s}(t)\mathbf{e}%
(t)+\mathbf{f}(t),\quad \mathbf{M=}\left[ 
\begin{array}{ccc}
0 & G & 0 \\ 
\alpha G & 0 & 0 \\ 
0 & 0 & 0%
\end{array}%
\right] ,  \label{tra}
\end{equation}%
where $\mathbf{e}(t)=(e_{1}(t),e_{2}(t),e_{3}(t))^T$ is a unit vector pointing in the instantaneous
swimming direction of the particle, $U_{s}(t)$ the
magnitude of the instantaneous swimming velocity along $\mathbf{e}(t)$ (in
other words, $\mathbf{U}_{s}(t)=U_{s}(t)\mathbf{e}(t)$), and $\mathbf{f}%
\equiv R_{U}^{-1}\widetilde{\mathbf{f}}$. Similarly, the director vector, $\mathbf{e}$,
follows the dynamics \citep{cof}%
\begin{equation}
\frac{\mathrm{d}{\mathbf{e}}}{\mathrm{d}t}=\left[ \omega _{\alpha}\mathbf{k+g%
}(t)\right] \mathbf{\times e}(t),  \label{rot}
\end{equation}%
where $\mathbf{g}\equiv R_{\Omega }^{-1}\widetilde{\mathbf{g}}$.

In the stochastic system of equations (\ref{tra})-(\ref{rot}), the equation
for the particle orientation, (\ref{rot}), can be solved first and its
solution can then be used in (\ref{tra}) to obtain the particle position. In
order to determine all components of the symmetric mean-square displacement
tensor, $\langle \mathbf{x}(t)\mathbf{x}(t)^{T}\rangle $, we
therefore have to compute first all components of the orientation
correlation matrix. %%%%%%

%%%%%%%%%%%%%%%%%%%%%%%%%
\section{Rotational probability distribution function and orientation
correlations}
\label{S:rot}

The orientation correlation matrix,  $\left\langle \mathbf{e}(t)%
\mathbf{e}(0)^{T}\right\rangle $, can be  evaluated if we know the
orientation transition probability distribution function (pdf),
 $P(\mathbf{e},t|\mathbf{e}_{0},0)$,  with $\mathbf{e}(0)\equiv \mathbf{e}_{0}$, governing the swimmer orientation, $\mathbf{e}(t)$. 
 Since the angular velocity of the spherical swimmer is along the k-axis, to determine $P$ we apply to equation~(\ref{rot}), a rotational transformation around the $%
\mathbf{k}$-direction of the frame fixed at the particle center, namely 
\begin{equation*}
\mathbf{e}(t)=\mathbb{R}(t)\mathbf{e}^{\prime }(t),\text{ \ }\mathbb{R}(t)%
\mathbf{=}\left[ 
\begin{array}{ccc}
\cos \omega _{\alpha }t & -\sin \omega _{\alpha }t & 0 \\ 
\sin \omega _{\alpha }t & \cos \omega _{\alpha }t & 0 \\ 
0 & 0 & 1%
\end{array}%
\right] ,
\end{equation*}%
where $\mathbf{e}^{\prime }(t)$ is the orientation vector in a co-ordinate system rotating with the flow vorticity. This transformation reduces
equation (\ref{rot}) to $\mathrm{d}{\mathbf{e}^{\prime }}/\mathrm{d}t=%
\mathbf{g}^{\prime }(t)\mathbf{\times e}^{\prime }(t),$ whose pdf for the
director vector is classically given by an infinite sum over spherical harmonics %
\citep{berne}. The transformation between the fixed and rotating frames of reference, in a spherical coordinate system with its polar axis along the ambient vorticity, only involves the two azimuthal angles($\varphi ^{\prime }=\varphi -\omega _{\alpha }t$ , $\varphi$ and $\varphi^{\prime}$ are respectively the azimuthal angles for fixed and rotating frames of references). Substituting this transformation, we find the required pdf of the director, $P$, in a general
linear flow as 
\begin{equation}
P(\mathbf{e},t|\mathbf{e}_{0},0)=\sum_{l=0}^{\infty }\sum_{m=-l}^{l}e^{-D_{\Omega
}l(l+1)t}Y_{l}^{m\ast }\left( \theta _{0},\varphi _{0}\right)
Y_{l}^{m}\left( \theta ,\varphi \right) e^{-im\omega _{\alpha }t},
\label{np}
\end{equation}%
where $\{Y_{l}^{m}\}$ are the spherical harmonics \citep{abramowitz}, $\{Y_{l}^{m\ast }\}$
their complex conjugates, and $\theta _{0}$ and $\varphi _{0}$ are the polar and azimuthal angles
for $ \mathbf{e}_{0}.$ %In equation \eqref{np}, $D_{\Omega }$ is the
In equation \eqref{np}, $D_{\Omega}$ is the rotary diffusivity for the particle. When it has a thermal origin, it is determined in terms of the amplitude of the Brownian force correlation (see equation \eqref{equationbrownian}), and is given by $ k_{B}T/R_\Omega$. The underlying random fluctuations in orientation may not be Brownian, however, in which case $D_\Omega$ may be directly inferred from the observed rate of change of the mean square angular displacement \citep{berg93}.
With the  explicit expression for the pdf  known,
the correlation matrix for the swimming orientation may then be evaluated. The $ij^{th}$ component is given by
\begin{equation}
\left\langle e_{i}(t)e_{j}(0)\right\rangle =\int \mathrm{d}^{2}e_{0}\int 
\mathrm{d}^{2}e\,e_{i}(t)e_{j}(0)G(\mathbf{e},t;\mathbf{e}_{0},0),  \label{1}
\end{equation}%
where $i,j$ are in \{1,2,3\}, and where $G(\mathbf{e},t;\mathbf{e}_{0},0)$
is the joint probability distribution function for the director vector with
orientation $\mathbf{e}_{0}$ at time $t=0$ and orientation $\mathbf{e}(t)$
at time $t$ \citep{berne}. For an assumed isotropic distribution of orientation at the initial instant, this joint probability is given by the product of the uniform pdf for $\mathbf{e}_{0}$ ($1/4\pi$) with the transition pdf ($P$) for the  orientation vector ${\bf e}(t)$, given that we know that the orientation was  $\mathbf{e}_{0}$  at $t=0$ and thus we have 
\begin{equation}
G(\mathbf{e},t;\mathbf{e}_{0},0)=\frac{1}{4\pi}P(\mathbf{e},t|\mathbf{e}_{0},0).
\end{equation}

Using this formalism, all components of $\left\langle \mathbf{e}(t)\mathbf{e}%
(0)^{T}\right\rangle $ may be systematically obtained. For example for $i=1$
and $j=2$, solving equation~(\ref{1}) directly leads to 
\begin{equation}
\left\langle e_{1}(t)e_{2}(0)\right\rangle =\frac{1}{4\pi }%
\sum_{l=0}^{\infty }\sum_{m=-l}^{l}e^{-D_{\Omega }l(l+1)t}e^{-im\omega
_{\alpha }t}D_{l}^{m},  \label{r2}
\end{equation}%
where%
\begin{eqnarray}
D_{l}^{m} &=&\left( -1\right) ^{m}\int \!\!\!\int h_{1}\mathrm{d}\theta _{0}%
\mathrm{d}\varphi _{0}\int \!\!\!\int h_{2}\mathrm{d}\theta \mathrm{d}%
\varphi ,  \label{d} \\
h_{1} &=&\sin ^{2}\theta _{0}\sin \varphi _{0}Y_{l}^{-m}\left( \theta
_{0},\varphi _{0}\right) ,\,\,h_{2}=\sin ^{2}\theta \cos \varphi
Y_{l}^{m}(\theta ,\varphi ).
\end{eqnarray}%
\newline
By orthogonality, we can show that $D_{l}^{m}=0$ if $l\neq 1$, and by
explicitly evaluating the coefficients $D_{1}^{m}$ we get 
\begin{equation}
\left\langle e_{1}(t)e_{2}(0)\right\rangle =-\frac{1}{3}e^{-2D_{\Omega
}t}\sin \omega _{\alpha }t.  \label{e12r}
\end{equation}
All other components of the orientation correlation matrix, $\langle \mathbf{%
e}(t)\mathbf{e}(0)^{T}\rangle $, can be similarly obtained, leading to the
final result 
\begin{equation}
\left\langle \mathbf{e}(t)\mathbf{e}(0)^{T}\right\rangle =\frac{1}{3}{%
e^{-2D_{\Omega }t}}\left[ 
\begin{array}{ccc}
\cos \omega _{\alpha }t & -\sin \omega _{\alpha }t & 0 \\ 
\sin \omega _{\alpha }t & \cos \omega _{\alpha }t & 0 \\ 
0 & 0 & 1%
\end{array}%
\right] .  \label{ee}
\end{equation}%
In the plane of the  linear flow, the components of the orientation
correlation matrix follow an exponential decay modulated by a harmonic
function with frequency equal to the linear flow-induced rotation rate. Note that upon 
setting $\omega _{\alpha }=0$ in equation \eqref{ee}, we recover the classical exponential decay in
orientation direction from Brownian motion in the absence of flow, $\langle {e}_{i}(t){e_{i}(0)}%
\rangle =e^{-2D_{\Omega }t}$ \citep{edwards}. 
%%%%%%%%%%%%%%%%%%%%%%%%%%%

%%%%%%%%%%%%%%%%%%%%%%%%%
\section{Mean-square displacement tensor}
\label{general}

We now turn to determining the general formula for the mean-square
displacement dyadic, i.e. the symmetric tensor $\langle \mathbf{x}(t)\mathbf{%
x}(t)^{T}\rangle $. An integration of equation (\ref{tra}) with initial
condition $\mathbf{x}(0)=0$ leads to the formal solution
\begin{equation}
\mathbf{x}(t)=\int_{0}^{t}U_{s}(t^{\prime })e^{\mathbf{M}(t-t')}%e^{-\mathbf{M}%t^{\prime }}
\mathbf{e}(t^{\prime })\mathrm{d}t^{\prime }+\int_{0}^{t}e^{%
\mathbf{M}(t-t')}%e^{-\mathbf{M}t^{\prime }}
\mathbf{f}(t^{\prime })\mathrm{d}%
t^{\prime }\mathbf{.}  \label{4}
\end{equation}%
Using the definition of the exponential matrix, one  can show that 
\begin{equation}
e^{\mathbf{M}(t-t')}%e^{-\mathbf{M}t^{\prime }}
=\left[ 
\begin{array}{ccc}
\cosh \left[ \sqrt{\alpha }G(t-t^{\prime })\right]  &\displaystyle \frac{1}{\sqrt{\alpha }%
}\sinh \left[ \sqrt{\alpha }G(t-t^{\prime })\right]  & 0 \\ 
\sqrt{\alpha }\sinh \left[ \sqrt{\alpha }G(t-t^{\prime })\right]  & \cosh 
\left[ \sqrt{\alpha }G(t-t^{\prime })\right]  & 0 \\ 
0 & 0 & 1%
\end{array}%
\right] 
\equiv \left[ 
\begin{array}{ccc}
b_{11} & b_{12} & 0 \\ 
b_{21} & b_{22} & 0 \\ 
0 & 0 & 1%
\end{array}%
\right] 
.  \label{4.1}
\end{equation}

We start by computing the diagonal elements of $\langle \mathbf{x}(t)\mathbf{%
x}(t)^{T}\rangle $. In order to do so we remark that, if $\beta $ denotes
one component of the particle position, $\beta =x_{i}$, then 
\begin{equation}
\frac{\mathrm{d}\left\langle \beta (t)\beta (t)\right\rangle }{\mathrm{d}t}%
=2\left\langle \beta \frac{\mathrm{d}{\beta }}{\mathrm{d}t}\right\rangle
\cdot  \label{diff}
\end{equation}%
With the initial condition $\beta (0)=0$, equation \eqref{diff} can be
integrated once to  obtain exactly 
\begin{equation}
\left\langle \beta (t)\beta (t)\right\rangle =2\int_{0}^{t}\left\langle
\beta \frac{\mathrm{d}{\beta }}{\mathrm{d}t}\right\rangle \mathrm{d}t.
\label{formula}
\end{equation}
We then proceed to perform the multiplications on the right-hand side of %
\eqref{diff} applied to each of the three components of $\mathbf{x}(t)$ given by equations (\ref{4}) and (\ref{4.1}).
After using the fluctuation-dissipation theorem stating that $\langle 
{{f}}_{i}(t){{f}_{j}}(t^{\prime })\rangle =2D_{B}\delta
(t-t^{\prime }),$ where $D_{B}$ is the Brownian diffusion constant, $%
D_{B}=k_{B}T/R_{U},$ and using that the random force and swimming direction
are not correlated we obtain 
\begin{eqnarray}
\left\langle x_{1}(t)\frac{\mathrm{d}{x_{1}}}{\mathrm{d}t}(t)\right\rangle 
&=&G\int_{0}^{t}U_{s}(t^{\prime })b_{1k}(t,t^{\prime
})\int_{0}^{t}U_{s}(t_{2})b_{2l}(t,t_{2})\left\langle e_{k}(t^{\prime
})e_{l}(t_{2})\right\rangle \mathrm{d}t_{2}\mathrm{d}t^{\prime }  \notag \\
&&+G\int_{0}^{t}b_{1l}(t,t^{\prime })\int_{0}^{t}b_{2k}(t,t_{2})\left\langle
f_{l}(t^{\prime })f_{k}(t_{2})\right\rangle \mathrm{d}t_{2}\mathrm{d}%
t^{\prime }  \notag \\
&&+U_{s}(t)\int_{0}^{t}U_{s}(t_{2})b_{1l}(t,t_{2})\left\langle
e_{1}(t)e_{l}(t_{2})\right\rangle \mathrm{d}t_{2}+D_{B},  \label{xx} \\
\left\langle x_{2}(t)\frac{\mathrm{d}{x_{2}}}{\mathrm{d}t}(t)\right\rangle 
&=&\alpha G\int_{0}^{t}U_{s}(t^{\prime })b_{1k}(t,t^{\prime
})\int_{0}^{t}U_{s}(t_{2})b_{2l}(t,t_{2})\left\langle e_{k}(t^{\prime
})e_{l}(t_{2})\right\rangle \mathrm{d}t_{2}\mathrm{d}t^{\prime }  \notag \\
&&+\alpha G\int_{0}^{t}b_{1l}(t,t^{\prime
})\int_{0}^{t}b_{2k}(t,t_{2})\left\langle f_{l}(t^{\prime
})f_{k}(t_{2})\right\rangle \mathrm{d}t_{2}\mathrm{d}t^{\prime }  \notag \\
&&+U_{s}(t)\int_{0}^{t}U_{s}(t_{2})b_{2l}(t,t_{2})\left\langle
e_{2}(t)e_{l}(t_{2})\right\rangle \mathrm{d}t_{2}+D_{B},  \label{yy} \\
\left\langle x_{3}(t)\frac{\mathrm{d}{x_{3}}}{\mathrm{d}t}(t)\right\rangle 
&=&U_{s}(t)\int_{0}^{t}U_{s}(t^{\prime })\left\langle
e_{3}(t)e_{3}(t^{\prime })\right\rangle \mathrm{d}t^{\prime }+D_{B},
\label{zz}
\end{eqnarray}
where $k,l$ are in \{1,2\} (Einstein summation notation).

In order to  compute the off-diagonal elements of $\langle \mathbf{x}(t)\mathbf{x}%
(t)^{T}\rangle $ we directly use the integration in equations (\ref{4})-(\ref%
{4.1}) which provides each component, $\left( x_{1},x_{2},x_{3}\right) $, of
the particle trajectory. The ensemble average of the direct multiplication
of these components together with the fact that random force and swimming
direction are not correlated leads to the general results 
\begin{eqnarray}
\left\langle x_{1}(t)x_{2}(t)\right\rangle &=&\int_{0}^{t}U_{s}(t^{\prime
})b_{1k}(t,t^{\prime })\int_{0}^{t}U_{s}(t_{2})b_{2l}(t,t_{2})\left\langle
e_{k}(t^{\prime })e_{l}(t_{2})\right\rangle \mathrm{d}t_{2}\mathrm{d}t^{\prime } 
\notag \\
&&+\int_{0}^{t}b_{1l}(t,t^{\prime })\int_{0}^{t}b_{2k}(t,t_{2})\left\langle
f_{l}(t^{\prime })f_{k}(t_{2})\right\rangle \mathrm{d}t_{2}\mathrm{d}%
t^{\prime },  \label{xy} \\
\left\langle x_{1}(t)x_{3}(t)\right\rangle &=&0,  \label{xz1} \\
\left\langle x_{2}(t)x_{3}(t)\right\rangle &=&0.  \label{yz}
\end{eqnarray}%
Independently of its  swimming kinematics,  for an active particle immersed
in a two-dimensional linear flow, the correlations between the particle
components in the plane of the linear flow and perpendicular to it are zero.

%%%%%%%%%%%%%%%%%%%%%%%
\section{Application to steady swimming}
\label{stead}
In the previous section, the general formulae for each component of the mean-square displacement dyadic, $\langle \mathbf{x}(t)\mathbf{x}(t)^{T}\rangle $, were derived. The final results, although analytically explicit, can be quite involved if $U_{s}(t)$ is a complicated function of time. To get further insight into the impact of swimming on the effective particle dynamics, we
apply our framework to the case of an active particle swimming in a steady
fashion, i.e. $U_{s}(t)=U$, where $U$ is a constant speed. 

To illustrate how this assumption can be exploited, we consider equation (\ref{xy}) for the correlation in the cross terms of the active particle, $\left\langle x_{1}(t)x_{2}(t)\right\rangle$. When $U_s=U$, using the fact that
\begin{equation}
\int_{0}^{t}b_{1l}(t,t^{\prime })\int_{0}^{t}b_{2k}(t,t_{2})\left\langle
f_{l}(t^{\prime })f_{k}(t_{2})\right\rangle \mathrm{d}t_{2}\mathrm{d}%
t^{\prime }=2D_{B}\int_{0}^{t}b_{1l}(t,t^{\prime })b_{2l}(t,t^{\prime })%
\mathrm{d}t^{\prime },  \label{delta}
\end{equation}
 equation (\ref{xy}) becomes 
\begin{eqnarray}
\left\langle x_1(t)x_2(t)\right\rangle  &=&U^{2}\int_{0}^{t}b_{1k}(t,t^{\prime
})\int_{0}^{t}b_{2l}(t,t_{2})\left\langle e_{k}(t^{\prime
})e_{l}(t_{2})\right\rangle \mathrm{d}t_{2}\mathrm{d}t^{\prime }  \label{xyr} \\
&&+2D_{B}\int_{0}^{t}b_{1l}(t,t^{\prime })b_{2l}(t,t^{\prime })\mathrm{d}%
t^{\prime }. \notag
\end{eqnarray}%
Using equation \eqref{4.1}, one easily finds that
\begin{equation}
2D_{B}\int_{0}^{t}b_{1l}(t,t^{\prime })b_{2l}(t,t^{\prime })\mathrm{d}%
t^{\prime }=D_{B}\frac{\sinh^2 \left( \sqrt{\alpha }Gt\right) }{G}+D_{B}%
\frac{\sinh^2 \left( \sqrt{\alpha }Gt\right)}{\alpha G}\cdot \label{I0} 
\end{equation}
Furthermore, an inspection of equation (\ref{xyr})  shows that four integrals  (denoted $F_1$ to $F_4$) have to be evaluated,
namely
\begin{equation}
F_{1}=U^{2}\int_{0}^{t}b_{11}(t,t^{\prime
})\int_{0}^{t}b_{21}(t,t_{2})\left\langle e_{1}(t^{\prime
})e_{1}(t_{2})\right\rangle \mathrm{d}t_{2}\mathrm{d}t^{\prime },
\label{F10}
\end{equation}
\begin{equation}
F_{2}=U^{2}\int_{0}^{t}b_{11}(t,t^{\prime
})\int_{0}^{t}b_{22}(t,t_{2})\left\langle e_{1}(t^{\prime
})e_{2}(t_{2})\right\rangle \mathrm{d}t_{2}\mathrm{d}t^{\prime },  \label{F2}
\end{equation}
\begin{equation}
F_{3}=U^{2}\int_{0}^{t}b_{12}(t,t^{\prime
})\int_{0}^{t}b_{21}(t,t_{2})\left\langle e_{2}(t^{\prime
})e_{1}(t_{2})\right\rangle \mathrm{d}t_{2}\mathrm{d}t^{\prime },  \label{F3}
\end{equation}
\begin{equation}
F_{4}=U^{2}\int_{0}^{t}b_{12}(t,t^{\prime
})\int_{0}^{t}b_{22}(t,t_{2})\left\langle e_{2}(t^{\prime
})e_{2}(t_{2})\right\rangle \mathrm{d}t_{2}\mathrm{d}t^{\prime }.  \label{F4}
\end{equation}
In fact, one can  see from the general equations (\ref{xx})-(\ref{xy})
that the four integrals, $F_1$ to $F_4$,  together with the equality in equation (\ref{delta}), are
common to all the non-zero components  of the tensor $\langle \mathbf{x}(t)\mathbf{x}%
(t)^{T}\rangle$ (apart from $\langle x_{3}x_{3}\rangle$). Evaluating $F_1$ to $F_4$ will thus allow us to obtain   explicit expressions for all components of the 
mean-square displacement tensor. 

In order to  compute the first integral $F_{1}$, one has to
pay attention to the relative magnitude of $t_{2}$ and $t^{\prime }$. Let us
rewrite the first integral as 
\begin{equation}
F_{1}=U^{2}\int_{0}^{t}b_{11}(t,t^{\prime })\left[ \int_{0}^{t^{\prime
}}b_{21}(t,t_{2})\left\langle e_{1}(t^{\prime })e_{1}(t_{2})\right\rangle 
\mathrm{d}t_{2}+\int_{t^{\prime }}^{t}b_{21}(t,t_{2})\left\langle
e_{1}(t_{2})e_{1}(t^{\prime })\right\rangle \mathrm{d}t_{2}\right] \mathrm{d}%
t^{\prime },  \label{F11}
\end{equation}%
so that for the term in the bracket we have $t^{\prime }\geq t_{2}$ in
the first integral while $t_{2}\geq t^{\prime }$ in the second one.
Inserting from equation (\ref{4.1}) the corresponding  values of $b_{kl},$ and
substituting the appropriate orientation correlations from equation (\ref{ee}%
) into equation (\ref{F11}), and after performing the integrations we finally obtain 
\begin{equation}
F_{1}\sim \frac{U^{2}\sqrt{\alpha }}{3}\frac{%
A_{1}+A_{2}}{k_{\alpha }},  {\rm as}\,\, t\to\infty,
\label{RF1}
\end{equation}%
where
\begin{eqnarray}
A_{1} &=& \frac{\left( a_{\alpha }^{2}-b_{\alpha
}^{2}+d_{\alpha }^{2}-c_{\alpha }^{2}\right) \cosh \left( \sqrt{\alpha }Gt%
\right) \sinh \left( \sqrt{\alpha }Gt\right) }{k_{\alpha }}  \label{A1}, \\
A_{2} &=&-2b_{\alpha } \frac{\sinh^2 \left( \sqrt{%
\alpha }Gt\right)}{2\sqrt{\alpha }G},  \label{C1}
\end{eqnarray}
with the constants $a_{\alpha },b_{\alpha },c_{\alpha },d_{\alpha }$ and $k_{\alpha }$ defined as%
\begin{eqnarray}
a_{\alpha } &=&\sqrt{\alpha }G\left( -4D_{\Omega }^{2}+G^{2}\alpha +\omega
_{\alpha }^{2}\right) ,  \label{c1} \\
b_{\alpha } &=&-8D_{\Omega }^{3}+2D_{\Omega }G^{2}\alpha -2D_{\Omega }\omega
_{\alpha }^{2},  \label{c2} \\
c_{\alpha } &=&4\sqrt{\alpha }GD_{\Omega }\omega _{\alpha },  \label{c3} \\
d_{\alpha } &=&4D_{\Omega }^{2}\omega _{\alpha }+G^{2}\alpha \omega _{\alpha
}+\omega _{\alpha }^{3},  \label{c4} \\
k_{\alpha } &=&16D_{\Omega }^{4}-8D_{\Omega }^{2}\left( G^{2}\alpha -\omega
_{\alpha }^{2}\right) +\left( G^{2}\alpha +\omega _{\alpha }^{2}\right) ^{2}.
\label{c5}
\end{eqnarray}
With an identical mathematical procedure, we can solve for the other three
integral terms  namely ($F_{2},F_{3}$ and $F_{4}$). For  $F_2$ we find 
\begin{equation}
F_{2} \sim \frac{U^{2}}{3}\frac{B_{1}+B_{2}+B_{3}}{%
k_{\alpha }},  {\rm as}\,\, t\to\infty,\label{RF2}
\end{equation}
where%
\begin{eqnarray}
B_{1} &=&\frac{2a_{\alpha }c_{\alpha }\sinh ^{2}%
\left( \sqrt{\alpha }Gt\right) -2b_{\alpha }d_{\alpha }\cosh ^{2}\left( 
\sqrt{\alpha }Gt\right) }{k_{\alpha }},  \label{A2} \\
B_{2} &= &-4D_{\Omega }\omega _{\alpha } \sinh ^{2}
\left( \sqrt{\alpha }Gt\right),\quad B_{3}=\frac{2d_{\alpha
}b_{\alpha }}{k_{\alpha }}\cdot \label{C2}
\end{eqnarray}
For $F_{3}$ we get%
\begin{equation}
F_{3}\sim \frac{\sqrt{\alpha }U^{2}}{3}\frac{%
C_{1}+C_{2}+C_{3}}{k_{\alpha }},    {\rm as}\,\, t\to\infty,
\label{RF3}
\end{equation}
where%
\begin{eqnarray}
C_{1} &= &\frac{4D_{\Omega }\omega _{\alpha
}}{\sqrt{\alpha }} \sinh^2 \left( \sqrt{\alpha }Gt\right),\quad C_{2}=\frac{1}{\sqrt{\alpha }}\frac{2a_{\alpha }c_{\alpha }}{k_{\alpha }%
},  \label{A3} \\
C_{3} &= &\frac{1}{\sqrt{\alpha }} \frac{2b_{\alpha
}d_{\alpha }\sinh ^{2}\left(\sqrt{\alpha }Gt\right) -2a_{\alpha }c_{\alpha
}\cosh ^{2}\left( \sqrt{\alpha }Gt\right) }{k_{\alpha }},  \label{C3}
\end{eqnarray}
and finally for $F_{4}$ we obtain%
\begin{equation}
F_{4} \sim \frac{U^{2}}{3}\frac{D_{1}+D_{2}}{k_{\alpha }},   {\rm as}\,\, t\to\infty,
\label{RF4}
\end{equation}
where%
\begin{eqnarray}
D_{1} &= &\frac{-2b_{\alpha }}{\sqrt{\alpha }} 
\frac{\sinh^2 \left( \sqrt{\alpha }Gt\right)}{2\sqrt{\alpha }G},\text{ }
\label{A4} \\
D_{2}&= &\frac{1}{\sqrt{\alpha }} \frac{%
\left( a_{\alpha }^{2}-b_{\alpha }^{2}+d_{\alpha }^{2}-c_{\alpha
}^{2}\right) \sinh \left( \sqrt{\alpha }Gt\right) \cosh \left( \sqrt{\alpha }%
Gt\right) }{k_{\alpha }}\cdot  \label{C4}
\end{eqnarray}

In order to compute the diagonal terms in $\langle \mathbf{x}(t)\mathbf{x}(t)^{T}\rangle $, we have five remaining integrals to calculate in equations (\ref{xx})-(%
\ref{zz}), which are constants and we have
\begin{eqnarray}
\lim_{t\rightarrow \infty }U^{2}\int_{0}^{t}b_{11}(t,t_{2})\left\langle
e_{1}(t)e_{1}(t_{2})\right\rangle \mathrm{d}t_{2} &=&-\frac{U^{2}}{3}\frac{%
b_{\alpha }}{k_{\alpha }},  \label{I1} \\
\lim_{t\rightarrow \infty }U^{2}\int_{0}^{t}b_{12}(t,t_{2})\left\langle
e_{1}(t)e_{2}(t_{2})\right\rangle \mathrm{d}t_{2} &=&-\frac{U^{2}}{3\sqrt{%
\alpha }}\frac{c_{\alpha }}{k_{\alpha }},  \label{I2} \\
\lim_{t\rightarrow \infty }U^{2}\int_{0}^{t}b_{21}(t,t_{2})\left\langle
e_{2}(t)e_{1}(t_{2})\right\rangle \mathrm{d}t_{2} &=&\frac{U^{2}}{3}\frac{%
\sqrt{\alpha }c_{\alpha }}{k_{\alpha }},  \label{I3} \\
\lim_{t\rightarrow \infty }U^{2}\int_{0}^{t}b_{21}(t,t_{2})\left\langle
e_{2}(t)e_{2}(t^{\prime })\right\rangle \mathrm{d}t_{2} &=&-\frac{U^{2}}{3}%
\frac{b_{\alpha }}{k_{\alpha }},  \label{I4} \\
\lim_{t\rightarrow \infty }U^{2}\int_{0}^{t}\left\langle
e_{3}(t)e_{3}(t^{\prime })\right\rangle \mathrm{d}t^{\prime } &=&\frac{U^{2}%
}{6D_{\Omega }}\cdot \label{I5}
\end{eqnarray}
Note that when $\alpha <0$, we have neglected all exponentially decaying terms in the equations above. In contrast, for $\alpha >0$,  we have neglected (and therefore omitted) terms of the form $e^{\sqrt{\alpha }Gt}e^{-2D_{\Omega }Gt}$ compared with those scaling as $e^{2\sqrt{\alpha }Gt}$ in the $A-D$ constants, and  as a result,   terms such as $B_{3}$ or $C_{2}$, or all constants in equations (\ref{I1})-(\ref{I4}),  can  also be neglected for $
\alpha >0$.

%%%%%%%%%%%%%
\section{Steady swimming in three different linear flows}
\label{3cases}

We computed so far the long-time components of the mean-square  displacement tensor, $\langle \mathbf{x}(t)\mathbf{x}(t)^{T}\rangle $, for a particle performing steady swimming in a general two-dimensional linear flow (arbitrary value of $\alpha $). In this section we apply our general results to the canonical cases of a solid-body rotation  ($\alpha =-1$), a simple shear flow ($\alpha =0$),  and a pure extension ($\alpha =1$). An important dimensionless number which will appear compares  two relevant time scales. One time scale is $D_{\Omega}^{-1}$, corresponding to the re-orientation of the swimmer due to rotary diffusion (thermal or otherwise), and the other time scale is $G^{-1}$, a characteristic time scale for the linear flow. The ratio between the two is a rotary P\'eclet number, $\Pe$, defined as  $\Pe\equiv G/(4D_\Omega)$ (the coefficient 4 is for mathematical convenience).  Swimmers with $\Pe \ll 1$ will  primarily be affected by the non-hydrodynamic fluctuating forces (responsible for rotary diffusion), whereas when $\Pe \gg 1$ we expect the external flow to play an important role.

\subsection{Solid-body rotation}
In this section we assume the external flow is a solid-body rotation. We then substitute equations (\ref{delta})-(\ref{I5}) into (\ref{xx})-(\ref{xy}), and evaluate these
components at $\alpha =-1$. After elementary  simplifications and by integrating 
equation (\ref{formula}), we obtain the analytical expressions for the long-time components of the mean-square displacement tensor as
\begin{eqnarray}
\left\langle x_{1}(t)x_{1}(t)\right\rangle  &=&{\left( \frac{U^{2}}{%
3D_{\Omega }}+2D_{B}\right) t,}  \label{xxpaperm1}
\\
\left\langle x_{3}(t)x_{3}(t)\right\rangle  &=&\left\langle x_{2}(t)x_{2}(t)\right\rangle  = \left\langle x_{1}(t)x_{1}(t)\right\rangle   \label{yypaperm1},\\
\left\langle x_{1}(t)x_{2}(t)\right\rangle  &=&0.  \label{xypaperm1} 
\end{eqnarray}
This result is, surprisingly, the same as the classical result for swimming-induced enhanced effective  diffusion \citep{lovely75,berg93}. 
Furthermore, if we chose $U=0$
into equations (\ref{xxpaperm1})-(\ref{xypaperm1}), one recovers the
classical result of a Brownian passive particle under an external flow
performing pure rotation \citep{miguel,foister} as
\begin{eqnarray}
\left\langle x_{1}(t)x_{1}(t)\right\rangle  &=&\left\langle
x_{2}(t)x_{2}(t)\right\rangle =\left\langle
x_{3}(t)x_{3}(t)\right\rangle=2D_{B}t,  \label{cla1} \\
\left\langle x_{1}(t)x_{2}(t)\right\rangle  &=&0.  \label{cla2}
\end{eqnarray}
The fact that this result is identical to the case without any external flow will be addressed in detail in  \S\ref{discussion}.

%%%%%%%%%%%%

\subsection{Simple shear flow}
We now turn to the case of a simple shear flow, for which $\alpha =0$. Exploiting the results from equations (\ref{delta})-(\ref{I5}) to evaluate equations (\ref{xx})-(\ref{xy}) at $\alpha =0$, together with  equation (\ref{formula}), gives us
the explicit analytical expressions for the long time components of
the tensor $\langle \mathbf{x}(t)\mathbf{x}(t)^{T}\rangle $, namely 
\begin{eqnarray}
\left\langle x_{1}(t)x_{1}(t)\right\rangle  &=&\left[ \frac{32\Pe^{2}D_{\Omega }^{2}D_{B}}{3}+\frac{16U^{2}D_{\Omega }}{9}\frac{\Pe^{2}}{1+\Pe^{2}}\right] t^{3}+\frac{4U^{2}}{3}\left[ \frac{%
\Pe^{4}-\Pe^{2}}{\left( 1+\Pe ^{2}\right) ^{2}}\right]
t^{2}  \notag \\
&&+\left[ \frac{4U^{2}}{3D_{\Omega }}\frac{\Pe^{2}}{\left( 1+\Pe^{2}\right) ^{2}}+\frac{U^{2}}{3D_{\Omega }\left( 1+
\Pe^{2}\right) }+2D_{B}\right] t,  \label{first}\\
\left\langle x_{2}(t)x_{2}(t)\right\rangle &=&\left[ \frac{U^{2}}{3D_{\Omega
}\left( 1+\Pe^{2}\right) }+2D_{B}\right] t,  \label{yyr}\\
\left\langle x_{3}(t)x_{3}(t)\right\rangle &=&\left[ \frac{U^{2}}{3D_{\Omega }}%
+2D_{B}\right] t,  \label{zzR}\\
\left\langle x_{1}(t)x_{2}(t)\right\rangle &=&\left[ 4D_{\Omega }D_{B}\Pe+\frac{2U^{2}}{3}\frac{\Pe}{1+\Pe^{2}}\right] t^{2}+%
\frac{U^{2}}{3D_{\Omega }}\left[ \frac{\Pe^{3}-\Pe}{\left(
1+\Pe^{2}\right) ^{2}}\right] t,  \label{xya0}
\end{eqnarray}
with the P\'eclet number, $\Pe$, defined above.  If we set $U=0$ in equations \eqref{first}-%
\eqref{xya0} then our results reduce to those known for Brownian motion of passive particles in
simple shear \citep{miguel,foister}. We obtain 
\begin{eqnarray}
\left\langle x_{1}(t)x_{1}(t)\right\rangle  &=&\displaystyle\frac{2}{3}%
G^{2}D_{B}t^{3}+2D_{B}t, \\
\left\langle x_{2}(t)x_{2}(t)\right\rangle  &=&\left\langle
x_{3}(t)x_{3}(t)\right\rangle =2D_{B}t,  \label{nou} \\
\left\langle x_{1}(t)x_{2}(t)\right\rangle  &=&GD_{B}t^{2}.  \label{nou2}
\end{eqnarray}%

The dynamics quantified by equations (\ref{first})-(\ref{xya0}), which combines
self-propulsion, Brownian motion, and an external simple shear flow, has a few
notable features. The diagonal component in the direction of the applied
simple shear flow, $\langle x_{1}x_{1}\rangle $, is dominated, at long time, by the 
$O(t^{3})$ superdiffusive scaling, with a coefficient enhanced, by the presence
of swimming, above its value for passive particles. The $\langle
x_{1}x_{1}\rangle $ component also includes an $O(t)$ diffusive term, which
was present for passive particles but is here enhanced by swimming, and a
new intermediate $O(t^{2})$ term. In contrast, the diagonal components in
the directions perpendicular to the shear flow, $\langle x_{2}x_{2}\rangle $
and $\langle x_{3}x_{3}\rangle$, grow linearly with time in an anisotropic
fashion. The effective diffusion constant in the shear direction, $\langle
x_{2}x_{2}\rangle $, is always smaller than that in the vorticity direction, 
$\langle x_{3}x_{3}\rangle $, due to shear-induced particle rotation. In
both cases, swimming increases the effective diffusion constant above the
purely Brownian diffusion constant for passive particles. Finally, as was
the case for passive Brownian motion, a non-zero cross correlation in
displacements in the plane of the flow also arises due to shear, $\langle
x_{1}x_{2}\rangle $, scaling quadratically in time, and enhanced by the
presence of swimming also leads to a new $O(t)$ term.

\subsection{Pure extension}
The final case we analyse is that of an active particle swimming steadily in a pure extensional  (irrotational)  flow. Following the analysis in the previous sections we now find the long-time components of $\langle \mathbf{x}(t)\mathbf{x}(t)^{T}\rangle $ to be 
given by
\begin{eqnarray}
\left\langle x_{1}(t)x_{1}(t)\right\rangle  &=&\frac{U^{2}}{48D_{\Omega }^2}
\left[ \frac{1}{\Pe\left(1+ 2\Pe \right) }\right] e^{2Gt}+\frac{D_{B}}{8D_{\Omega }\Pe}e^{2Gt},  \label{xxpaper1} \\
\left\langle x_{2}(t)x_{2}(t)\right\rangle  &=&
\left\langle x_{1}(t)x_{2}(t)\right\rangle  = \left\langle x_{1}(t)x_{1}(t)\right\rangle,\\
\left\langle x_{3}(t)x_{3}(t)\right\rangle  &=&\left( \frac{U^{2}}{%
3D_{\Omega }}+2D_{B}\right) t.  \label{zzpaper1}
\end{eqnarray}%
Note that in order to derive the equations above we have neglected all algebraic terms which are subdominant compared to  $e^{2Gt}$ as long as $G\neq 0$. Once again, by setting $U=0$ into
equations (\ref{xxpaper1})-(\ref{zzpaper1}), one recovers (to within exponentially small corrections) the classical long-time correlations results of a Brownian passive particle in an extensional flow \citep{miguel,foister}%
\begin{equation}
\left\langle x_{1}(t)x_{1}(t)\right\rangle =\left\langle
x_{2}(t)x_{2}(t)\right\rangle =\left\langle x_{1}(t)x_{2}(t)\right\rangle
=D_{B}G^{-1}e^{2Gt}/2.  \label{pe}
\end{equation}%
The effect of activity is to lead to the same exponential scaling as for passive particle, but with an enhanced coefficient.

%%%%%%%%%%%
{\section{Extension to run-and-tumble swimmers}\label{Section_tumbling}

In this section, we extend the analysis to a spherical active particle whose orientation de-correlates due to stochastic instantaneous tumble events, in addition to the rotary diffusion process assumed above. The particle now `runs', on average, in a given direction during which its orientation evolves continuously
 due to rotary diffusion. However, such runs are interrupted by `tumbles' that lead to large impulsive changes in orientation. The statistics of the tumbles are  well approximated by a Poisson process for the bacterium \textit{E.~coli} \citep{berg93}. The duration of a run, $t_{run}$, is therefore governed by an exponential distribution function,  $ e^{-t_{run}/\tau}/\tau$, where  $\tau^{-1}$ is the average tumbling frequency.  
 
 In addition to describing the temporal statistics of tumbling events, one has to provide  a model for the correlations between the  pre- and post-tumble orientations. For instance in \textit{E.~coli}, an average angular change of $68^\circ$ per tumble has been observed \citep{
berg93}, indicative of a positive correlation. The original transition probability distribution introduced in  \S \ref{S:rot}, $P(\mathbf{e},t | \mathbf{e_{0}},0)$,  is again transformed to a coordinate system rotating with the particle ($\mathbf{e}\rightarrow \mathbb{R}(t)\mathbf{e}^{\prime }$, 
the rotation matrix $\mathbb{R}(t)$ being defined in \S\ref{S:rot}), and $P(\mathbf{e}',t | \mathbf{e_{0}},0)$  now satisfies the equation 
\begin{equation}
 \frac{\partial{P}}{\partial{t}}- D_{\Omega}\nabla^{2}_{\mathbf{e'}}P+\frac{1}{\tau}\left(P-\int K(\mathbf{e'}|\mathbf{e''})P(\mathbf{e''},t|\mathbf{e_0},0)\d \mathbf{e''}\right)=\delta(\mathbf{e'}-\mathbf{e_0})\delta(t)\label{Equation_Tumble_1},
\end{equation}
where $\nabla_{\mathbf{e'}}$ is the gradient operator over the unit sphere
\citep{othmer1988,SubKoch}. The exponential distribution of run lengths ensures that the probability of a tumble occurring in an infinitesimal interval \textit{\d t} remains the same ($\propto  \d t/\tau$), independent of any earlier tumbling events. As described in equation~\eqref{Equation_Tumble_1}, tumbling may be regarded as a linear collision process with `direct' (third term on the left-hand side of equation \ref{Equation_Tumble_1}) and `inverse' events (fourth term), which lead, respectively, to a decrease and an increase in the probability density in the differential angular interval ($\mathbf{e} ,\mathbf{e}+d\mathbf{e}$). The kernel,  ${K}(\mathbf{e'}|\mathbf{e''})$, is the transition probability density associated with a tumble from $\mathbf{e''}$ to $\mathbf{e'}$, which in the absence of chemical gradient  is expected to be a function of $\mathbf{e'\cdot e''}$ only. 
Conservation of probability further requires that $\int K(\mathbf{e'}|\mathbf{e''})\d \mathbf{e'}=\int K(\mathbf{e'} |\mathbf{e''})\d \mathbf{e''}=1$. An example of a kernel  satisfying the above constraints is 
\begin{equation}
K(\mathbf{e'}|\mathbf{e''})=\frac{\beta}{(4\pi \sinh \beta)} \exp({\beta \mathbf{e'\cdot e''}}),\label{equation_kernel}
\end{equation}
where tuning the parameter $\beta$ allows for a wide range of correlations \citep{SubKoch}. For $\beta\rightarrow 0$, $K=1/4\pi$,  corresponding to perfectly random tumbles (and an average angular change of $90^{0}$), while for $\beta\rightarrow \infty $, there is only an infinitesimally small change in orientation, and thence, a near balance between the direct and inverse collision terms. The value $\beta=1$ leads to an average angle change, during tumbles, close to that observed for \textit{E.~coli}.

Interestingly,  in the limit $\beta\rightarrow\infty$ and $\tau\rightarrow 0$, and with $\beta \tau$ finite, the combination of the direct and inverse collision terms in equation (\ref{Equation_Tumble_1}) simplifies to the orientational Laplacian multiplied by a factor proportional to $(\beta \tau)^{-1}$. The simplification may be seen by noting that, for $\beta \rightarrow \infty$, tumbles are increasingly local events in orientation space, and accordingly, $P(\mathbf{e''},t|\mathbf{e}_{0},0)$ in the inverse collision term may be expanded about $P(\mathbf{e'},t|\mathbf{e}_{0},0)$ as a Taylor series, leading to the orientational Laplacian at the leading order. In this limit, the governing equation, equation~(\ref{Equation_Tumble_1}), again describes orientational de-correlation due to a rotary diffusion process, but with the rotary diffusivity being now given by the sum of the original rotary diffusivity, $D_{\Omega}$, and the added contribution of O$(\beta\tau)^{-1}$ from small-amplitude tumbles.

We solve equation (\ref{Equation_Tumble_1})  by expanding the orientation probability distribution in terms
of the surface spherical harmonics,  $Y_l^m (\theta',\varphi')$, defined in  \S\ref{S:rot}. The kernel, on account of its dependence on the scalar argument $ \mathbf{e'\cdot e''}$ alone, can be expanded in terms of Legendre polynomials in $ \mathbf{e' \cdot e''}$. Thus we formally get 
\begin{equation}
P(\mathbf{e'},t|\mathbf{e_0},0)=\displaystyle\sum\limits_{l=0}^\infty\sum\limits_{m=-l}^{m=l} c_{lm} Y_l^m(\theta',\varphi') g_{lm}(t),\end{equation}
and
\begin{equation}\label{KK}
  \textit{K}(\mathbf{e'}|\mathbf{e''})= \displaystyle\sum\limits_{n=0}^\infty a_n P_n(\mathbf{e'\cdot e''}),
  \end{equation}
 where $\{a_n\}$, $\{c_{lm}\}$ are constants, the $\{g_{lm}(t)\}$ are functions of time and $P_n$ refers to the Legendre polynomial of degree \textit{n}, which may be expressed in terms of the original spherical harmonics by means of the addition theorem \citep{abramowitz}. The solution can be then transformed back to a space-fixed coordinate system and is finally given by
\begin{equation}
 P(\mathbf{e},t|\mathbf{e_0},0)=\displaystyle\sum \limits_{l=0}^\infty\sum\limits_{m=-l}^{m=l}Y_{l}^{m}(\theta,\varphi)Y_{l}^{m*}(\theta_0,\varphi_0)e^{-\left[D_{\Omega}l(l+1)+\frac{1}{\tau}-\frac{4\pi a_l}{(2l+1)\tau}\right]t}e^{-i m \omega_{\alpha}t}\label{Equation_Tumble_2},
\end{equation}
where $Y_{l}^{m}$ and $Y_{l}^{m*}$ are defined in \S\ref{S:rot}. From  equation~(\ref{Equation_Tumble_2}), it is seen that the relaxation of the initial delta function in orientation space to an isotropic distribution is characterized by a denumerable infinity of decaying exponentials. In the absence of rotary diffusion, and with the additional simplification of the tumbles being perfectly random (i.e.~\textit{K}($\mathbf{e'}|\mathbf{e''}$) = 1/$4 \pi$),  equation (\ref{Equation_Tumble_2}) reduces to 
\begin{equation} P(\mathbf{e},t| \mathbf{e_{0}},0)=\frac{1}{4\pi}(1-e^{-\frac{t}{\tau}})+\delta(\mathbf{e}-\mathbf{e_0})e^{-\frac{t}{\tau}}, \label{Equation_Tumble_3}\end{equation}
where we have used $a_0 = 1/4\pi$, $D_{\Omega}=0$ and $\omega_{\alpha}=0$. The expression in (\ref{Equation_Tumble_3}) shows that in this limit, the  initial delta function in orientation space now relaxes to isotropy as a single exponential.

It is of interest to compare  equation~(\ref{Equation_Tumble_2}), that includes stochastic de-correlation due to both diffusion and tumbling, to  equation \eqref{np}, which  quantified only rotary diffusion. The introduction of tumbling only leads to a difference in the decay rates of the exponentials which now include an additional contribution proportional to 1/$\tau$. This is because the eigenfunctions in both cases are the surface spherical harmonics themselves, and the introduction of the tumbling terms only affects the distribution of eigenvalues.

With the probability distribution, $P(\mathbf{e},t | \mathbf{e_{0}},0)$, known from  equation (\ref{Equation_Tumble_2}) ,  the calculation from  \S \ref{S:rot} for the average orientation correlation matrix, ${\left\langle\mathbf{e}(t)\mathbf{e}(0)\right\rangle^T}$, can be carried out and we now obtain
\begin{equation}
\left\langle \mathbf{e}(t)\mathbf{e}(0)^{T}\right\rangle =\frac{1}{3}
{\exp\left\{{-\left(2D_{\Omega }+\frac{1}{\tau}-\frac{4 \pi a_{1}}{3\tau}\right)t}\right\}}\left[ 
\begin{array}{ccc}
\cos \omega _{\alpha }t & -\sin \omega _{\alpha }t & 0 \\ 
\sin \omega _{\alpha }t & \cos \omega _{\alpha }t & 0 \\ 
0 & 0 & 1%
\end{array}%
\right] 
,\label{Equation_Tumble_4}
\end{equation}
where $a_{1}$ is the coefficient of the first-order Legendre polynomial in the expansion of the tumbling kernel; for $K$ as in  equation~\eqref{equation_kernel} we have   $a_1$ = $(3\beta \cosh\beta-3 \sinh \beta)/(4\pi\beta \sinh\beta)$ and $a_1\approx 0.075$ for $\beta=1$. Note that in the limit $\beta\rightarrow \infty$, we obtain $a_1= 3/4\pi$, and equation (\ref{Equation_Tumble_4}) reduces to (\ref{ee}).

A comparison between the expressions in equations (\ref{ee}) and (\ref{Equation_Tumble_4}) reveals that the effect of correlated tumbling is to yield an effective rotary diffusivity that is larger than the true diffusivity by an amount $({1}/{2}-{2 \pi a_{1}}/{3})/{\tau}$, even though the actual de-correlation mechanism is, of course, no longer diffusive. All results obtained above in  \S\ref{3cases} for the three canonical flows with rotary diffusion alone, can thus be generalised to include also run-and-tumble dynamics by merely replacing the rotary diffusivity, $ D_{\Omega}  $, by an effective diffusion constant, denoted $\tilde D_{\Omega}  $, and given by
\begin{equation}
\label{newD}
\tilde D_{\Omega}  =D_{\Omega }+\frac{1}{\tau }\left(\frac{1}{2}-\frac{2 \pi a_{1}}{3}\right)\cdot
\end{equation}
Note that this effective rotary diffusivity may also be arrived at by noting that the total rate of decorrelation due to independent stochastic processes must be the sum of the individual decorrelation rates. The individual decorrelation rates due to rotary diffusion and tumbling may be obtained from the respective translational diffusivities, $D = U^2/6D_{\Omega}$ for rotary diffusion vs.~$D = [U^2/(3-4\pi a_{1})]\tau$ for tumbling alone, implying that the total rate of decorrelation must involve the combination $D_{\Omega} + (3-4\pi a_{1})/(6\tau)$.

\section{Discussion}\label{discussion}
In the cases of simple shear and extensional flow, we saw that the activity of the particles leads to long-time temporal scalings for the tensor $\langle {\bf x} {\bf x} \rangle$ similar to those obtained for the dynamics of passive particles, albeit with increased coefficients. In this section we examine the order of magnitude of our results, investigate the physical origin of the scalings obtained, estimate the typical time scale after which the enhancement is observed, and discuss the  relevance  of our results for biology and bioengineering.

\subsection{Enhanced mean-square displacement}
We first  summarise the results from \S\ref{3cases} and \S\ref{Section_tumbling} in table \ref{tab1}. For all three flows, we show  the terms dominating the behaviour at $t\to \infty$ and separate the passive ($U=0$) case from the case %where the particle swims, rotary diffuses and  ($U \neq 0$, $D_\Omega$ finite, $\tau \rightarrow \infty$)%
where particle executes a run-and-tumble motion with rotary diffusion during the runs ($U \neq 0$). The results for the active swimmer with rotary diffusion alone may be obtained by formally replacing the effective diffusion constant,  $\tilde{D}_\Omega$, by the true rotary diffusivity, ${D}_\Omega$. In all cases, the strength of the flow is characterised by the rotary P\'eclet number, $\Pe$, the ratio of the time scale characterising the intrinsic orientation de-correlation due both to rotary diffusion and tumbling and a characteristic flow time scale.

\[\begin{array}{l|cccc}
\toprule
&&\alpha =-1 &\alpha =0 &\alpha =1  \\
&&\rm{Rotation}&\rm{Simple \,\,shear}& \rm{Extension}\\
\hline 
\left\langle x_{1}x_{1}\right\rangle\quad\quad&U=0& 2D_Bt &
\displaystyle \frac{2}{3}G^2{D_{B}} t^{3}
& \displaystyle\frac{D_{B}}{2G}e^{2Gt} \\ \\
&  U\neq 0  & \displaystyle  +\frac{U^2}{3 \tilde{D}_\Omega}t &+\displaystyle \frac{G^2 U^{2}}{9\tilde{D}_\Omega(1+
\Pe^{2})}
 t^{3}
& +  \displaystyle\frac{U^{2}}{12 G \tilde{D}_\Omega\left(1+ 2\Pe \right)}e^{2Gt} \\
\hline
\left\langle x_{2}x_{2}\right\rangle&U=0&2D_Bt&2D_Bt& \displaystyle\frac{D_{B}}{2G}e^{2Gt} \\ \\
&U\neq 0&\displaystyle+\frac{U^2}{3 \tilde{D}_\Omega }t &
\displaystyle + \frac{U^{2}}{3 \tilde{D}_\Omega\left( 1+\Pe^{2}\right) } t
 & +  \displaystyle\frac{U^{2}}{12 G \tilde{D}_\Omega \left(1+ 2\Pe\right)}e^{2Gt} \\
\hline
\left\langle x_{3}x_{3}\right\rangle&U=0&2D_Bt&2D_Bt&2D_Bt \\ \\
&U \neq 0&\displaystyle+\frac{U^2}{3 \tilde{D}_\Omega}t&\displaystyle+\frac{U^2}{3 \tilde{D}_\Omega}t&\displaystyle
+\frac{U^2}{3D_\Omega}t \\
\hline
\left\langle x_{1}x_{2}\right\rangle&U=0& 0&G D_{B} t^2 &\displaystyle\frac{D_{B}}{2G}e^{2Gt} \\ \\
&U\neq 0&0&\displaystyle+\frac{U^{2} G}{6 \tilde{D}_\Omega \left(
1+\Pe^{2}\right) } t^{2}  & 
+  \displaystyle\frac{U^{2}}{12 G \tilde{D}_\Omega \left(1+ 2\Pe\right)}e^{2Gt}  \\
\bottomrule
\end{array}\]
\captionof{table}{\small Long-time  components of the mean-square  displacement tensor, $\langle \mathbf{x}(t)\mathbf{x}(t)^{T}\rangle 
$, for three different linear flows, namely 
rotation ($\alpha=-1$)
shear ($\alpha=0$) and extension ($\alpha=1$). In each row, the results first show the dynamics in the no-swimming case ($U=0$) followed by  the additional term due to activity ($U\neq 0$). Recall that 
we have defined the P\'eclet number as $\Pe=G/\tilde{D}_\Omega$ where the effective rotational diffusivity, $\tilde{D}_\Omega$, is given in equation \eqref{newD}.}
\label{tab1}
\bigskip

In the limit $\Pe \ll 1$, for all cases in table \ref{tab1}, the ratio between the mean-square displacement in active (random tumbling) and the passive case is given by
\begin{equation}\label{ratio1}
 \frac{\langle x_ix_j\rangle_{U\neq 0}}{ \langle x_ix_j\rangle_{U=0}} \sim \frac{U^2}{\tilde D _\Omega D_B}\cdot
\end{equation}
From equation~\eqref{ratio1} we see that the flow strength, $G$, has disappeared, and the effect of the activity is of the same order as the ratio between the typical swimming-induced translational diffusivity in the absence of external flow, 
$ U^2/\tilde D_\Omega$,  and the Brownian diffusivity, $D_B$. Note also that since the linear flow is two-dimensional, the scaling in \eqref{ratio1} remains actually valid for all values of $\Pe$ in the case of   $\langle x_{3}x_{3}\rangle$.

In the case of strong flows, $\Pe\gg 1$, and from table \ref{tab1} we obtain the ratio of mean squared displacements for the active and passive cases as: 
\begin{equation}\label{ratio2}
\frac{\langle x_ix_j\rangle_{U\neq 0}}{ \langle x_ix_j\rangle_{U=0}}\sim \frac{U^2}{\tilde D _\Omega D_B
\Pe^n}, 
\end{equation}
where $n=2$ for simple shear, $n=1$ in the case of extensional flow and $n=0$ for solid-body rotation. Thus, for simple shear and extensional flow, the strong flow limit leads to a relative decrease of the contribution from the particle activity. For solid-body rotation, however, the mean-squared displacement is the same as that known for a swimmer in a quiescent fluid medium. This can be  seen from a reference frame which is rotating with the flow wherein the only orientation decorrelation mechanism for an active particle is rotary diffusion and potentially tumbling (see below for a further discussion).

\subsection{Physical scalings}
\label{scalings}
One may use simple physical arguments to recover the scalings seen in table \ref{tab1}. The arguments presented below are for particles without tumbling, and the generalization to include run-and-tumble dynamics, as indicated above, may be done by way of an effective rotary diffusivity.  

We begin by recalling that, in a quiescent fluid, the characteristic step size scales as $U/D_\Omega$, the decorrelation time scales as $1/D_{\Omega}$, leading to a translational diffusivity scaling as $U^2/D_\Omega$, and thus $\langle x^2\rangle  \sim (U^2/D_\Omega) t$. This may now be used to obtain the convectively enhanced scalings for the mean-square displacements in simple shear and extensional flow. For pure shear and in the weak flow limit,  diffusion along the gradient direction leads to $ x_2\sim [(U^2/D_\Omega) t]^{1/2}$, and the corresponding distance traversed along the flow direction is $x_1\sim G x_2 t$, implying that $\langle x_1x_1\rangle  \sim O((G U)^2 t^3/D_{\Omega})$.  In the strong flow limit, the characteristic step size in the gradient direction is $U/G$, since the displacement due to swimming is cut off by the rotation due to the ambient vorticity.  The decorrelation time scales as  $1/D_{\Omega}$ , leading to a flow-dependent translational diffusivity of  $(U/G)^2 D_\Omega$ and $\langle x_2x_2 \rangle \sim (U/G)^2 D_\Omega t$. In turn, this implies that $\langle x_1x_1 \rangle \sim O({(G x_2 t)}^2) \sim U^2 D_\Omega t^3$, which is the limiting form, for high $\Pe$, of the results for simple shear flow in table \ref{tab1}.

In the case of extensional flow, the deterministic terms imply that $x_{1} \sim e^{Gt}$, and thus $\langle x_{1}x_{1}\rangle  \sim  e^{2Gt}-1$, for a swimmer starting from the origin. The prefactor in $\langle x_{1}x_{1} \rangle $, given by  $U^2/(GD_\Omega)$, is obtained by Taylor expansion by noting that for times of order $D_\Omega^{-1}$ (much smaller than $G^{-1}$ in the weak flow limit), $
\langle x_{1}x_{1}\rangle $ must still be diffusive. In the strong flow limit, the prefactor scales as  $U^2/G^2$, and is thus  independent of $D_\Omega$. In this limit, the decorrelation due to rotary diffusion occurs at a much larger time compared to the flow time scale,  and there is thus a direct transition from the short-time ballistic regime to the exponential enhancement driven by the ambient flow. 

\subsection{The peculiar case of solid body rotation}

It is of interest to note that diffusivity in solid body rotation is unaffected by vorticity strength, whereas in simple shear flow, the diffusivity in the gradient direction decreases with flow strength as $\propto G^{-2}$ as shown by the above scaling arguments. The orbital frequency (time taken to complete an entire circuit along a closed streamline) and the rotation frequency (equal to half the ambient vorticity) are exactly the same for an active particle in solid-body rotation, and this leads to the lack of dependence on the flow vorticity. Solid-body rotation is thus a singular limit. For the family of elliptic linear flows, with  $\alpha =-|\alpha|$, that span the interval between simple shear and solid-body rotation, there is always a mismatch between the orbital frequency, $G\sqrt{|\alpha|}$, and the rotation frequency, $ G(1+|\alpha|)/2$. This mismatch leads to a finite displacement in the deterministic limit. An active swimmer in an elliptic linear flow ends up swimming indefinitely, and with a periodic reversal in direction, within a region whose spatial extent is $\sim U/[G(1-\sqrt{|\alpha|})]$.  The reversal in direction happens on a time scale of order $G(1-\sqrt{|\alpha|})^{-1}$, and thus, the behaviour of the mean square displacement in an elliptic linear {flow} depends on the relative magnitudes of the intrinsic decorrelation time,  $D_\Omega^{-1}$, and the aforementioned deterministic reversal time. When $D_\Omega^{-1} \ll G(1-\sqrt{|\alpha|})^{-1}$, then the long-time diffusivities along the principal axes of the elliptical streamlines are independent of the flow strength; note that this is the only limit relevant to solid-body rotation. In the strong flow limit, however, we have $D_\Omega^{-1} \gg G(1-\sqrt{|\alpha|})^{-1}$, and the diffusivities scale as $\propto G^{-2}$. This, and the additional dependence on $|\alpha|$, may be obtained by noting that the characteristic step size is now of order $U/[G(1-\sqrt{|\alpha|})]$, while the decorrelation time is still $\sim D_\Omega^{-1}$. So, the long-time diffusivity (along the minor axis of the closed streamlines) scales as $(U/G(1-\sqrt{|\alpha|})^2 D_{\Omega}$. The breakdown of this argument, and the flow-independence of the diffusivity for solid-body rotation, arises from the divergence of the elementary step size in the limit $\alpha \rightarrow - 1$.

\subsection{Time scales for enhancement}
Another issue of interest, in the case of shear flows, is the time one has to wait in order to observe the enhanced mean-square displacement, $\sim t^3$, along the flow direction, $\langle x_1 x_1\rangle$. That time scale can be  obtained by comparing the order of magnitudes of the $\sim t^2$ and $\sim t^3$ terms in equation \eqref{first}. For a weak shear flow, $\Pe\ll  1$, we get a cross-over at a critical time scale such that $\tilde D_\Omega t\sim U^2 /(\tilde D_\Omega D_B + U^2)$. Assuming that activity leads to enhanced mean-square displacement, we thus have $U^2\gg \tilde  D_\Omega D_B$ (see equation \ref{ratio1}), and therefore see that the cross over occurs  on the order of the rotational  time scale, $t\sim  \tilde D_\Omega^{-1} $. In the case of a strong shear flow, $\Pe\gg 1$, we get that the cross over occurs when  $t\sim U^2/(U^2\tilde D_\Omega + G^2D_B)$. If we assume again to be in the enhanced regime, corresponding to $U^2\gg \tilde D_\Omega D_B G^2 \tau^2$ (see equation \ref{ratio2}), and thus $U^2\tilde D_\Omega \gg  G^2D_B$, leading again to $t \sim \tilde D_\Omega^{-1}$. The relevant time to obtain the enhanced mean square displacement is therefore independent on weak vs.~strong nature of the flow, and is always the typical orientation decorrelation time. A similar analysis can be carried out for the cross term, $\langle x_1 x_2\rangle $, with similar results.

 \subsection{Relevance to biology and bioengineering}

From a practical standpoint, when can we expect these results to be quantitatively  important? Let us consider a small biological or synthetic  swimmer with a typical size of 1 $\mu$m.  At room temperature and in water this leads to a Brownian diffusion constant of  $D_B\approx0.22$ $\mu$m$^2$s$^{-1}$ and $D_\Omega \approx 0.16$ s$^{-1}$ leading to a thermal time scale of $\approx$ 3s. The estimate in equation \eqref{ratio1} says that, for weak flows,  the critical swimming speed to observe an enhancement is $U_c\sim (D_\Omega D_B)^{1/2}\approx$ 200 nm\,s$^{-1}$. Micron-sized swimmers both biological \citep{lp09} and synthetic \citep{chemical2} typically go much faster than this value, and thus the effect quantified here should result in enhancement by orders of magnitude and  easily seen experimentally. 

In the presence of a strong flow, the critical swimming speed necessary in order to  observe an enhanced mean-square motion is increased due to the  $\Pe ^n$ term in equation \eqref{ratio2}. What is the typical value of a deformation rate, $G$, in a  practical situation? We consider two cases. The first is that of planktonic bacteria \citep{guasto12}, which are subject to wind-driven flows with rms deformation rates of up to $G\sim10$ s$^{-1}$  on the smallest length scales \citep{jimenez1997oceanic}. These {flows} typically {possess} both extensional ($n=1$)   and viscous ($n=2$) components and are typically turbulent, but given that the Kolmogorov length scale is at least a few millimeters, they appear laminar on the scale of a micron-size organism.  In that case, the critical velocity becomes $U_c\sim (D_\Omega D_B)^{1/2}(\Pe)^{n/2}\sim$~1~$ \mu$m\,s$^{-1}$ for extensional flow and $U_c\sim$~5~$\mu$m\,s$^{-1}$ for shear and rotation. These swimming speeds are  below typical velocities in biological locomotion,  and thus the random motion of bacteria in oceanic flow is expected to be strongly affected  by their activity.
 
A second situation of interest would be that of synthetic swimmers in the blood flow, where in this case the motion is dominated by shear  ($n=2$). In large  blood vessels we have $G\sim 10^2$ s$^{-1}$ \citep{pedley1980fluid}, leading to a critical swimming speed for enhanced motion in a shear flow of $U_c\sim$ 50 $\mu$m\,s$^{-1}$, on the upper limit of the synthetic swimming speeds measured in the laboratory.    In contrast, for flow in capillaries we have much larger deformation rates, up to $G\sim10^4$ s$^{-1}$ \citep{lipowsky1978distribution}, leading to a large value $U_c\sim$ 5 mm\,s$^{-1}$. Whereas the random motion of small synthetic swimmers is expected to be affected by both blood flow and the swimmer motion in large vessels, the effect of swimming in small capillaries will probably be negligible. 

\subsection{Summary and perspective}

In summary we have addressed theoretically the stochastic dynamics of spherical active particles diffusing  in an incompressible, two-dimensional linear flow. After deriving the general framework valid for an arbitrary time-dependent swimming velocity of the particles, we focused on the special case of steadily swimming particles and, have illustrated our analytical results on three different flows: solid-body rotation, simple shear, and extension. We have also shown  that the results can be extended to a particle which executes a run-and-tumble motion, as a model for the dynamics of bacteria. Compared to passive colloidal particles, we have shown that the activity of the particle leads to the same long-time scalings but with increased values of the coefficients,   which can be physically rationalized (see summary in  table~\ref{tab1}). 
By comparing the  new terms with those obtained for passive particles we have shown that the activity of the particles could lead to enhancement by orders of magnitude of their mean-square displacement, for example for planktonic bacteria subject to oceanic turbulence.  Our results could thus be further exploited to quantify the ability of specific small-scale biological organisms to sample their surroundings.

The calculations in the paper were made under a number of assumptions which suggest ways in which the study could be generalized. We have assumed the flows to be of an infinite extent, whereas for example in a biological setting it is clear that the presence of boundaries would play an important role. We have also assumed the active particle to be spherical, allowing us to perform all calculations analytically.  For non spherical bodies, relevant for example for elongated bacteria,  equations \eqref{din} would include an additional  term which  depends on the symmetric part of the rate-of-strain tensor, and would require the use of numerical computations to derive the effective long-time dynamics of the active particle (or restriction of the analysis to certain asymptotic regimes in the rotary P\'eclet number). One important  difference between the dynamics of spherical and non-spherical particles is that whereas spherical particles undergo   uniform rotation at a rate proportional to the flow vorticity, non-spherical particles rotate along Jeffery orbits, and for large aspect ratios, end up spending a significant amount of time aligned in certain directions (the flow-vorticity plane in simple shear).

Finally, beyond thermal forces and run-and-tumble, other sources of directional change could be address with our modeling approach, in particular run-and-reverse for bacteria \citep{guasto12}, phase-slips in eukaryotic flagella \citep{Polin2009}, collisions \citep{ishi} or even non-thermal turbulent fluctuations in flow vorticity in environmental flows \citep{jimenez1997oceanic}. Despite these limitations, we hope that our study will provide new insight into the interplay between orientation decorrelation, external flows, and activity, and  will be valuable in order to develop coarse-grained theories of swimming populations in  complex, external flows.

\smallskip This work was funded in part by the Consejo Nacional de Ciencia y Tecnologia of Mexico (Conacyt postdoctoral fellowship to M. S.) and the US National Science Foundation (Grant CBET-0746285 to E.L.).
\bibliographystyle{jfm}
\bibliography{shear}

\end{document}